# Time-varying k-domain modulation around a point sink in time reversal cavity

Xin Liu[1#], Jin-Shan Cui[1#], Ren Wang[*1], Yanhe Lv[2], and Bing-Zhong Wang[1]

[1]*Institute of Applied Physics, University of Electronic Science and Technology of China, Chengdu, 611731, China*
[2]*Department of Electrical and Computer Engineering, National University of Singapore, 117583, Singapore*

*# Xin Liu and Jin-Shan Cui contribute equally to the paper.*
*\*rwang@uestc.edu.cn*

**Abstract:** This paper derives and investigates a time-varying *k*-domain modulation in vector form using a time-reversal (TR) field decomposition theory proposed for the first time. First, the proposed theory illustrates that the TR field can exhibit super-resolution property from the perspective of spatial focusing pattern if a point sink is set at the initial source point. Afterward, the instantaneous TR fields with and without the point sink, as well as their *k*-domain patterns are compared to illustrate the time-varying *k*-domain modulation, which accounts for the super-resolution property. The phenomenon observed and derived in this paper shows great potential in the applications empowered by super-resolution focusing, such as wireless communication carrying the subwavelength information and high spatial resolution wireless power transfer.

## 1.  Introduction

Spatial wave number *k* modulation is a fundamental concept in optics and acoustics, playing a pivotal role in achieving super-resolution imaging and focusing beyond the classical diffraction limit [1-7]. Numerous specially engineered materials have been proposed to implement *k*-domain modulation, including waveguides [8, 9], grating plates [10], grid resistive films [11], hyperbolic lenses [12], nanowires [13], photonic crystals [14, 15], metalenses [16-19], and scattering media [20-22]. Beyond employing the above structurally engineered materials, Fink et al have also revealed in the realm of acoustics that converging waves, characterized by the scalar anti-causal Green's function, can exhibit high-*k* components around a point sink, thereby enabling super-resolution focusing [33-36]. These existing methods primarily utilize non-time-varying *k*-domain modulation for super-resolution imaging or focusing.

In recent years, time-varying metamaterials have garnered significant attention for their potential to manipulate electromagnetic waves [23-26]. Space-time metamaterials, in particular, offer the promise of achieving *k*-domain modulation [27-32] and hold the potential for time-varying *k*-domain modulation [23, 24].

By proposing a theory of TR field decomposition, this paper derives and analyzes, for the first time, the time-varying *k*-domain modulation in the vector form around a point sink, empowering the super-resolution focusing in optics. The super-resolution focusing within the established electromagnetic TR cavity model has been demonstrated with either a current element or a magnetic current element as the TR source. The theoretical findings of this paper hold significance for super-resolution optics and time-varying optics, opening new avenues for exploration and applications in the field.

## 2.  Super-resolution imaging from the perspective of spatial focusing pattern

TR field in frequency domain derived from TR cavity (TRC) theory [37] is analyzed and decomposed into the causal Green's function based and the anti-causal Green's function-based field components in this section. From the perspective of the spatial focusing pattern, it is revealed that the total TR field shows a focusing spot on the order of half a wavelength, while the isolated anti-causal Green's function-based field, forming by setting a perfect point sink at the position of the initial source, is beyond the diffraction limit with its focusing spot much smaller than half a wavelength. Assuming an infinitesimal unit electric current with current density $\boldsymbol{i} = (0,0,1)$, Fig. 1 illustrates the TR procedure and compares the focusing spots with and without the presence of the point sink.

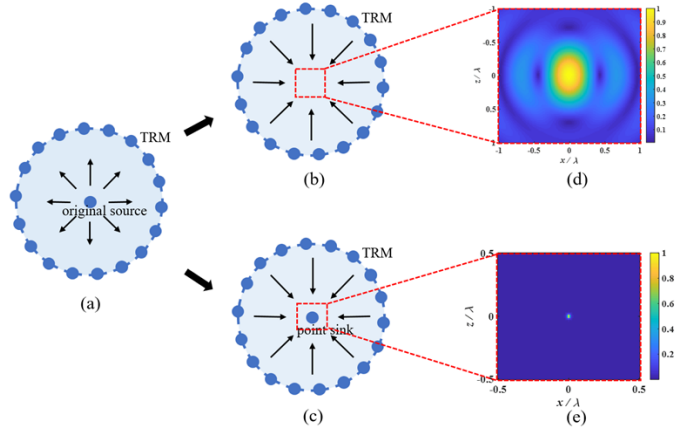

Fig. 1. TR procedure and the spatial focusing patterns with and without the point sink: (a) the signal is emitted by the initial source and reaches the TR mirrors (TRM), which are arranged in a circle around the source to form a TRC; the signals received by TRM are time-reversed and emitted in the case that (b) without the point sink ; (c) with a point sink at the initial source point; (d) normalized spatial focusing patterns in (b); (e) normalized spatial focusing patterns in (c).

## 2.1 Decomposition of the TR field

According to the TR cavity (TRC) theory [37], the frequency-domain TR field corresponding to an initial source of the electric current $\boldsymbol{J} = \delta(\boldsymbol{r}-\boldsymbol{r}_0)\boldsymbol{i}(\omega)$ can be expressed as [38]

$$\boldsymbol{E}^{TR}(\boldsymbol{r},\boldsymbol{r}_0;\omega) = j\omega\mu\left\{\overline{\overline{G}}_e^*(\boldsymbol{r},\boldsymbol{r}_0) - \overline{\overline{G}}_e(\boldsymbol{r},\boldsymbol{r}_0)\right\}\boldsymbol{i}^*(\omega) \tag{1}$$

where $\boldsymbol{r}$ and $\boldsymbol{r}_0$ are the locations of the observation point and the initial source point respectively, $\boldsymbol{i}$ is the current density, $\omega$ is the angular frequency, $\overline{\overline{G}}_e$ is the electric dyadic Green's function, and * indicates complex conjugate.

The dyadic Green's function appearing in Eq. (1) is related to the scalar Green's function by

$$\overline{\overline{G}}_e(\boldsymbol{r},\boldsymbol{r}_0) = \left(\overline{\overline{I}} + \frac{\nabla\nabla}{k^2}\right)G(\boldsymbol{r},\boldsymbol{r}_0) \tag{2}$$

where $\overline{\overline{I}}$ is the unit dyadic, $k$ is the wave phase constant. Although the TR cavity theory is of consideration, the scalar Green's function here can still be regarded as free-space Green's function, as Eq. (3), owing to the case that the distance between $\boldsymbol{r}$ and $\boldsymbol{r}_0$ is infinitely close and the scattered field in the vicinity of $\boldsymbol{r}_0$ is negligible.

$$G(\boldsymbol{r},\boldsymbol{r}_0) = \frac{\exp(-jk|\boldsymbol{r}-\boldsymbol{r}_0|)}{4\pi|\boldsymbol{r}-\boldsymbol{r}_0|} \tag{3}$$

Then, $\boldsymbol{E}^{TR}(\boldsymbol{r},\boldsymbol{r}_0;\omega)$ can be written as

$$\begin{aligned}
\boldsymbol{E}_{total}^{TR}(\boldsymbol{r},\boldsymbol{r}_0;\omega) &= \mathrm{j}\omega\mu\left\{\left(\bar{\bar{I}}+\frac{\nabla\nabla}{k^2}\right)\left(G^*(\boldsymbol{r},\boldsymbol{r}_0)-G(\boldsymbol{r},\boldsymbol{r}_0)\right)\right\}\boldsymbol{i}^*(\omega) \\
&= -\frac{\omega\mu}{\lambda}\left\{\left(\bar{\bar{I}}+\frac{\nabla\nabla}{k^2}\right)\left(\frac{\sin(k|\boldsymbol{r}-\boldsymbol{r}_0|)}{k|\boldsymbol{r}-\boldsymbol{r}_0|}\right)\right\}\boldsymbol{i}^*(\omega)
\end{aligned} \quad (4)$$

Eq. (4) reaches the extremum when $\boldsymbol{r}=\boldsymbol{r}_0$, revealing that the TR field can be spatially focused on the initial source point. The anti-causal Green's function $G^*(\boldsymbol{r},\boldsymbol{r}_0)$ and causal Green's function $G(\boldsymbol{r},\boldsymbol{r}_0)$ represent the converging wave and the diverging wave respectively, here $G^*(\boldsymbol{r},\boldsymbol{r}_0)$ is the conjugate of $G(\boldsymbol{r},\boldsymbol{r}_0)$. Thus, the TR focusing field can be regarded as the superposition of converging waves and diverging waves, with a distribution similar to that of standing waves. The first zero-crossing point of $\{\sin(k|\boldsymbol{r}-\boldsymbol{r}_0|)/(k|\boldsymbol{r}-\boldsymbol{r}_0|)\}\bar{\bar{I}}$ occurs at $k|\boldsymbol{r}-\boldsymbol{r}_0|=\pi$, the entire spherical surface with a radius of $\lambda/2$ from $\boldsymbol{r}_0$, $\lambda$ is the wavelength in the free space. Thus, the size of the focusing spot of the TR electromagnetic field is constrained by the diffraction limit.

However, if there is a perfect point sink at the position of the initial source during the process of TR focusing, the diverging wave represented by causal Green's function $G(\boldsymbol{r},\boldsymbol{r}_0)$ cannot be generated, resulting that the TR field is only determined by the converging wave represented by anti-causal Green's function $G^*(\boldsymbol{r},\boldsymbol{r}_0)$. The TR field in this case can be expressed as

$$\boldsymbol{E}_{anti-causal}^{TR}(\boldsymbol{r},\boldsymbol{r}_0;\omega) = \mathrm{j}\omega\mu\left\{\left(\bar{\bar{I}}+\frac{\nabla\nabla}{k^2}\right)G^*(\boldsymbol{r},\boldsymbol{r}_0)\right\}\boldsymbol{i}^*(\omega) \quad (5)$$

It can be derived from Eq. (3) that both the isolated converging and diverging wave distribute as an impulse function in the vicinity of the initial source, which means that the focusing spot of Eq. (5) has broken the diffraction barrier with its size much smaller than $\lambda/2$.

## 2.2 Spatial focusing pattern of the TR field without the point sink

### 2.2.1 TR field without the point sink with the electric current as the initial excitation source

The total TR field without the point sink depicted by Eq. (4) needs to be further expanded in the rectangular coordinate system to examine the spatial distribution.

Denote the distance from any point on the source to the observation point by $R$

$$R = |\boldsymbol{r}-\boldsymbol{r}_0| = \sqrt{(x-x_0)^2+(y-y_0)^2+(z-z_0)^2} \quad (6)$$

The total TR field under the excitation of the electric current source is derived:

$$\boldsymbol{E}_{total}^{TR}(R;\omega) = -\frac{\omega\mu}{\lambda}\left\{\begin{bmatrix}-\dfrac{\sin(kR)}{kR}-\dfrac{3\cos(kR)}{(kR)^2}+\dfrac{3\sin(kR)}{(kR)^3}\end{bmatrix}\boldsymbol{e}_R\boldsymbol{e}_R \\ +\begin{bmatrix}\dfrac{\sin(kR)}{kR}+\dfrac{\cos(kR)}{(kR)^2}-\dfrac{\sin(kR)}{(kR)^3}\end{bmatrix}\bar{\bar{I}}\right\}\boldsymbol{i}^*(\omega) \quad (7)$$

where $\boldsymbol{e}_R$ is the unit vector directed from the source point toward the observation point. $\boldsymbol{E}_{total}^{TR}(R;\omega)$ in Eq. (7) is a superposition of two vectors, one of which is along the direction of $\boldsymbol{e}_R$ and another is in the same direction as vector $\boldsymbol{i}$, the polarization of the source.

Assume that an infinitesimal unit electric current is centered at the origin and lies on the $z$-axis, with current density $\boldsymbol{i}=(0,0,1)$.

The amplitude distribution of the TR field in the *xOz* plane, parallel to the polarization of the excitation source, is shown in Fig. 2. It exhibits obvious spatial anisotropy, which is fundamentally different from the isotropy of the scalar TR field. From Fig. 2, it is obvious that in the *xOz* plane, the focusing radius along the *x*-axis is the shortest, while that along the *z*-axis is the longest. The focusing spot is on the order of half a wavelength, limited by the diffraction barrier.

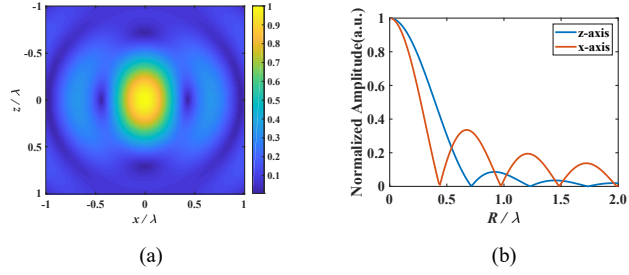

(a)           (b)

Fig. 2. Normalized amplitude distribution of the total TR field in the *xOz* plane corresponding to unit electric current: (a) spatial distribution; (b) amplitude distribution along *x*- and *z*- axes.

### 2.2.2 TR field without the point sink with the magnetic current as the initial excitation source

When the magnetic current is the initial excitation source, the expression for the TR field can be obtained in a similar way as above, except that a different type of dyadic Green's function is utilized.

$$\boldsymbol{E}^{TR}(\boldsymbol{r},\boldsymbol{r}_0;\omega) = \left\{ \overline{\overline{G_m}}(\boldsymbol{r},\boldsymbol{r}_0) - \overline{\overline{G_m}}^*(\boldsymbol{r},\boldsymbol{r}_0) \right\} \boldsymbol{i}^{m*}(\omega) \tag{8}$$

where $\boldsymbol{i}^m$ is the magnetic current source, $\overline{\overline{G_m}}$ is the magnetic dyadic Green's function and can be obtained by the relation

$$\overline{\overline{G_m}}(\boldsymbol{r},\boldsymbol{r}_0) = \nabla G(\boldsymbol{r},\boldsymbol{r}_0) \times \overline{\overline{I}} \tag{9}$$

Under the condition that the region to be investigated is free space or the distance between the observation point $\boldsymbol{r}$ and the initial excitation source point $\boldsymbol{r}_0$ is infinitely close, the Green's function in Eq. (9) can be replaced by the free-space Green's function in Eq. (3).

Then, the expression of the total TR electric field under the excitation of the magnetic current source can be obtained from Eq. (8) as

$$\boldsymbol{E}^{TR}_{total}(\boldsymbol{r},\boldsymbol{r}_0;\omega) = \nabla \left[ \frac{\text{jsin}(-k|\boldsymbol{r}-\boldsymbol{r}_0|)}{2\pi|\boldsymbol{r}-\boldsymbol{r}_0|} \right] \times \overline{\overline{I}} \cdot \boldsymbol{i}^m(\omega) \tag{10}$$

Denote the distance from any point on the source to the observation point by *R* in Eq. (6). Then it can be deduced from Eq. (8) that

$$\boldsymbol{E}^{TR}_{total}(R;\omega) = \frac{-\text{j}}{2\pi} \cdot \frac{kR\cos(kR)-\sin(kR)}{R^3} \begin{vmatrix} 0 & z_0-z & y-y_0 \\ z-z_0 & 0 & x_0-x \\ y_0-y & x-x_0 & 0 \end{vmatrix} \cdot \boldsymbol{i}^m(\omega) \tag{11}$$

Assuming that the initial excitation magnetic current source is a unit magnetic current source placed at the origin and along the *x*-direction, that is $\boldsymbol{i}^m = (1,0,0)$, its amplitude distribution along different directions in the *xOy* plane is examined and shown in Fig. 3. It can be seen from Fig. 3(a) that the TR field corresponding to the magnetic current source still exhibits anisotropic spatial distribution, but it is drastically different from the TR field corresponding to the electric current source.

The TR field amplitude is zero along the *x*-axis due to the placement of the magnetic current, and the focusing spot of the TR field is on the order of half a wavelength at line of $y = 0.04\lambda$. However, in the direction perpendicular to the polarization of the initial magnetic current source, the result is completely different from that of the electric current source, where the field amplitude is zero at the location of the initial source, rather than the maximum field amplitude. The size of the focusing spot is a little smaller than half a wavelength. The total TR field corresponding to a magnetic initial source is still constrained by the diffraction limit.

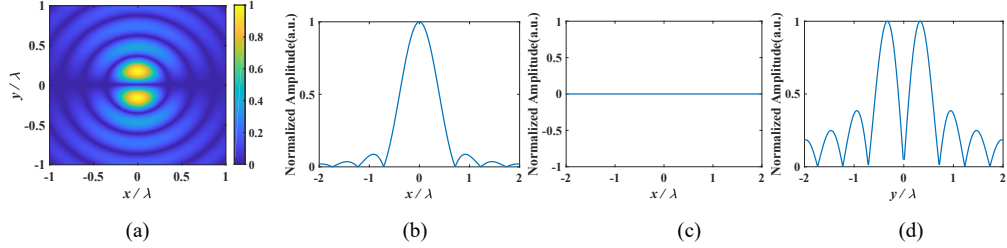

Fig. 3. Normalized amplitude distribution of the total TR field corresponding to unit magnetic current in the *xOy* plane: (a) spatial distribution; amplitude distribution along (b) $y = 0.04\lambda$; (c) $y = 0$; (d) $x = 0$.

## 2.3 Spatial focusing pattern of the TR field in the presence of a point sink

### 2.3.1 TR field in the presence of a point sink with the electric current as the initial excitation source

Expand Eq. (5) in the rectangular coordinate system, then the anti-causal Green's function-based TR field around a point sink can be expressed as

$$\boldsymbol{E}_{anti-causal}^{TR}(R;\omega) = -\frac{j\omega\mu}{2\lambda}\left\{\begin{bmatrix}-\frac{\exp(jkR)}{kR} - \frac{3j\exp(jkR)}{(kR)^2} + \frac{3\exp(jkR)}{(kR)^3}\end{bmatrix}\boldsymbol{e}_R\boldsymbol{e}_R \\ +\begin{bmatrix}\frac{\exp(jkR)}{kR} + \frac{j\exp(jkR)}{(kR)^2} - \frac{\exp(jkR)}{(kR)^3}\end{bmatrix}\overline{\overline{I}}\right\}\boldsymbol{i}^*(\omega) \quad (12)$$

For comparison, an electric current source located at the origin and polarized in z-direction, with current density $\boldsymbol{i} = (0,0,1)$, is also introduced. The amplitude distribution of the TR field in the *xOz* plane, parallel to the polarization of excitation source, is shown in Fig. 4. Different from Fig. 2(a), Fig. 4(a) shows a spatial distribution with an extremely small focusing spot at the origin. To investigate further, the TR field amplitude distribution along the *x*-direction and *z*-direction are displayed in Fig. 4(b), which both distribute as an impulse function near the position of the initial source. The radius of the focusing spot in this case is much smaller than half a wavelength, which means that the TR electromagnetic wave around a point sink exhibits the super-resolution focusing property.

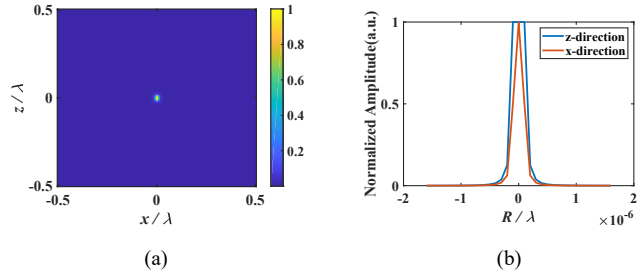

Fig. 4. Normalized amplitude distribution of the TR field around a point sink in the *xOz* plane corresponding to unit electric current: (a) spatial distribution; (b) amplitude distribution along *x*- and *z*-axis.

### 2.3.2 TR field in the presence of a point sink with the magnetic current as the initial excitation source

From Eqs. (8) and (9), the expression of the TR electric field under the excitation of the magnetic current source in the presence of a point sink can be obtained as follows.

$$\boldsymbol{E}^{TR}_{anti-causal}(R;\omega) = \frac{1}{2\lambda} \cdot \frac{(jkR-1)\exp(jkR)}{kR^3} \begin{vmatrix} 0 & z_0-z & y-y_0 \\ z-z_0 & 0 & x_0-x \\ y_0-y & x-x_0 & 0 \end{vmatrix} \cdot \boldsymbol{i}^m(\omega) \qquad (13)$$

The same assumption that the a unit magnetic current source placed at the origin and along the *x*-direction is applied, then the TR field can be calculated according to Eq. (13), and its amplitude distribution along different directions is examined separately and shown in Fig. 5.

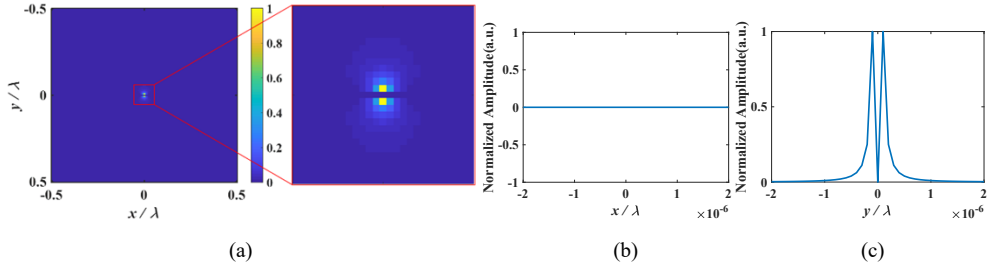

Fig. 5. Normalized amplitude distribution of the TR field around a point sink in the *xOy* plane corresponding to unit magnetic current: (a) spatial distribution; amplitude distribution along (b) *x*-axis; (c) *y*-axis.

It can be seen from Fig. 5(a) that the TR field corresponding to the magnetic current source with a point sink at the location of the initial source still shows a spatial distribution with two extremely small focusing spot at the origin. Furthermore, the TR field amplitude is also zero along the *x*-axis due to the placement of the magnetic current, while the pattern along the *y*-axis depicts two focusing spots, both much smaller than half a wavelength. Then, the TR electromagnetic wave corresponding to the magnetic current source also exhibits the super-resolution focusing property around a point sink.

In conclusion, the TR electromagnetic focusing will be not constrained by the diffraction limit in the presence of a point sink, whether the electric current or the magnetic current is applied as the initial excitation source. The convergent wave point sink put at the source position can eliminate the causal Green's function based divergent wave generated following the convergent wave, leading to the isolated convergent wave.

### 3. *K*-domain modulation behind the super-resolution property

The resolution limit of conventional far-field optical imaging is approximately half a wavelength, which can be accounted for by the loss of subwavelength features included in evanescent waves. The field distribution in a plane can be converted into its spatial spectrum, or *k*-domain pattern, using two-dimensional Fourier transform methods, each point there represents a plane wave with a particular spatial wavenumber. Plane waves with transverse wavenumbers larger than operational wavenumber, known as evanescent waves, decay exponentially with increasing propagation distance in the direction normal to the field plane. Thus, the subwavelength information will not reach the far-field region.

For the super-resolution property that the TR field around a point sink exhibits, it is revealed that evanescent waves with large transverse wavenumbers are spontaneously converted into propagating waves with small transverse wavenumbers to breaking the diffraction limit, and this *k*-domain modulation is observed to be time-varying.

The study of TR waves originated and flourished in the field of acoustics, where acoustic waves are scalar waves, while there are few reports on the instantaneous scalar TR field. The instantaneous scalar TR field and its k-domain patterns are elaborated in Section 1 in **Supplement 1**, and the time-varying video in this case can be found in **Visualization 1**.

*3.1 K-domain modulation with the electric current as the initial excitation source*

Following the assumption of the Sections 2.2.1 and 2.3.1, a unit electric current located at the origin and polarized in z-direction is also applied in the analysis below, where the k-domain patterns in the *xOz* plane that the source parallel to are examined.

Referring to Eq. (10) and Eq. (12), the instantaneous expressions of the vector TR field in the *xOz* plane with the electric current $\boldsymbol{i} = (0,0,1)$ as the initial excitation source are derived as follows respectively.

$$\boldsymbol{E}_{total}^{TR}(R;t) = \mathrm{Re}\left\{-\frac{\omega\mu}{\lambda}\left\{\left[-\frac{\sin kR}{kR} - \frac{3\cos kR}{k^2R^2} + \frac{3\sin kR}{k^3R^3}\right]\left(\frac{xz}{R^2}\boldsymbol{e}_x + \frac{z^2}{R^2}\boldsymbol{e}_z\right) + \left[\frac{\sin kR}{kR} + \frac{\cos kR}{k^2R^2} - \frac{\sin kR}{k^3R^3}\right]\boldsymbol{e}_z\right\}\exp(j\omega t)\right\} \quad (14)$$

$$\boldsymbol{E}_{anti-causal}^{TR}(R;t) = \mathrm{Re}\left\{\frac{j\omega\mu}{2\lambda}\left\{\left[-\frac{\exp(jkR)}{kR} - \frac{3j\exp(jkR)}{k^2R^2} + \frac{3\exp(jkR)}{k^3R^3}\right]\left(\frac{xz}{R^2}\boldsymbol{e}_x + \frac{z^2}{R^2}\boldsymbol{e}_z\right) + \left[\frac{\exp(jkR)}{kR} + \frac{j\exp(jkR)}{k^2R^2} - \frac{\exp(jkR)}{k^3R^3}\right]\boldsymbol{e}_z\right\}\exp(j\omega t)\right\} \quad (15)$$

where $\boldsymbol{e}_x$ and $\boldsymbol{e}_z$ are the unit vectors along the positive *x*- and *z*-direction, respectively. Re[·] satisfies the identity $\mathrm{Re}[\boldsymbol{E}\cdot\exp(j\omega t)] = [\boldsymbol{E}\cdot\exp(j\omega t) + \boldsymbol{E}^*\cdot\exp(-j\omega t)]/2$. Therefore, the TR fields in the *xOz* plane here consists of two components in the *x*- and *z*- directions, which will be investigated separately.

For the total TR field without the point sink indicated by Eq. (14), the k-domain patterns are independent of $\omega t$, thus the diagrams of *x*- and *z*- components when $\omega t = \pi / 2$ are shown in Fig. 6 and Fig. 7 for demonstration, respectively. The bright circle in Figs. 6(b) and 7(b) represents the spectral frequencies for which $k_x^2 + k_y^2 = k_0^2$, where $k_0 = 2\pi/\lambda_0$, $\lambda_0$ is the operating wavelength in vacuum, $k_x$ and $k_y$ are the wavenumbers along the *x*- and *y*- directions respectively. The components outside the circle with $k_x^2 + k_y^2 > k_0^2$ leading to the imaginary $k_z$ contribute to the evanescent waves which exponentially decay in the *z*-direction, while that within the circle contribute to the propagating waves. Despite the considerably differences in the field distributions and k-domain patterns in Fig. 6 and Fig. 7, the one thing they have in common is the non-zero components within the circle, which indicates that only propagating waves are present in both the *x*- and *z*- components.

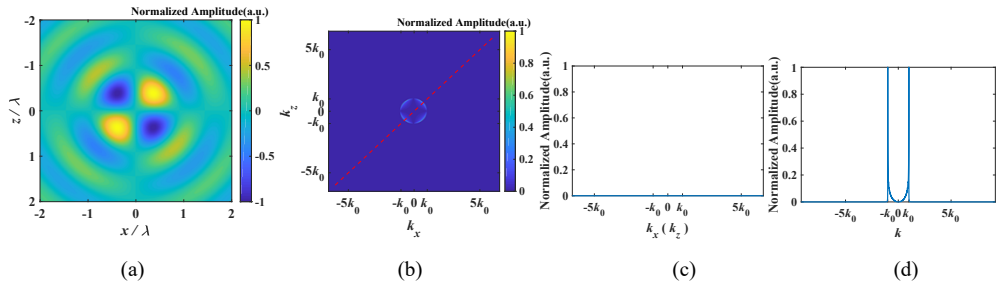

(a)　　　　　　　　(b)　　　　　　　　(c)　　　　　　　　(d)

Fig.6. Normalized field distribution and *k*-domain patterns for the *x*-component of the total TR field without the point sink in the *xOz* plane when $\omega t = \pi / 2$: (a) normalized spatial distribution; (b) normalized *k*-domain pattern of (a); normalized *k*-domain pattern (c) along the line of $k_z = 0$ or $k_x = 0$ and (d) along the red dashed line in (b).

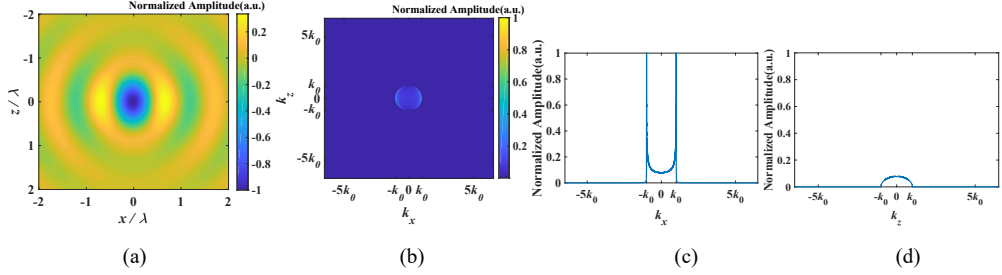

(a)  (b)  (c)  (d)

Fig.7. Normalized field distribution and *k*-domain patterns for the *z*-component of the total TR field without the point sink in the *xOz* plane when $\omega t = \pi / 2$: (a) normalized spatial distribution; (b) normalized *k*-domain pattern of (a); normalized *k*-domain pattern along the lines of (c) $k_z = 0$ and (d) $k_x = 0$.

As for the TR field around a point sink indicated by Eq. (15), the *k*-domain pattern is time-varying and the non-zero components here can be observed outside the circle. Fig. 8 shows the normalized field distributions and *k*-domain patterns of the *x*-component for three typical values of $\omega t$, meaning that there are almost only evanescent waves when $\omega t = m\pi + \pi / 2$ (*m* = 0, 1, 2….), only propagating waves when $\omega t = m\pi$, and both two types of waves when $\omega t$ is any other values, such as $\omega t = \pi / 5$.

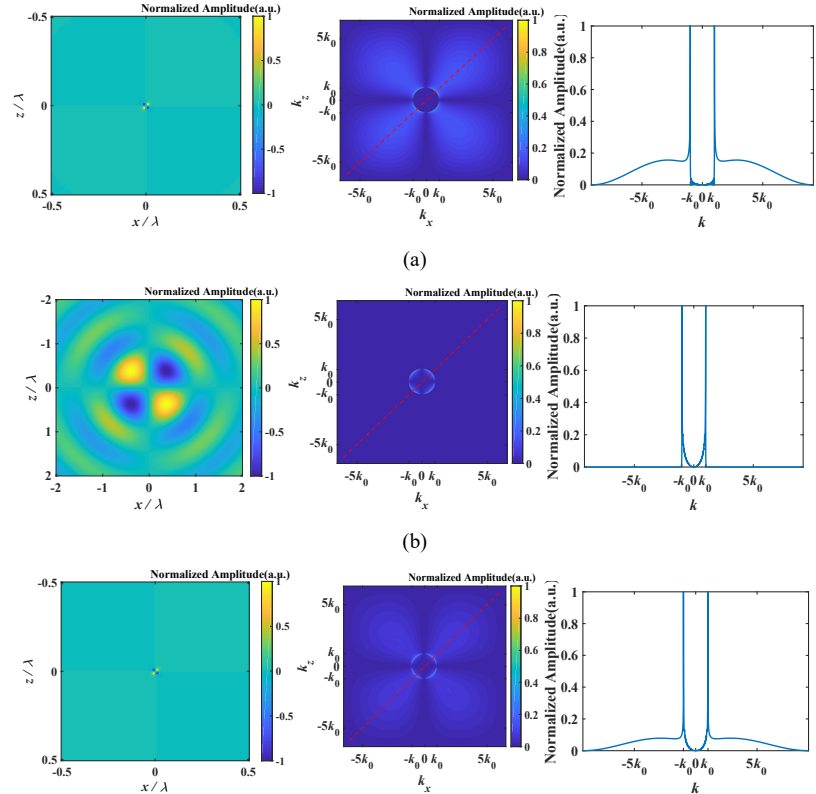

(a)

(b)

(c)

Fig. 8. Normalized field distributions and $k$-domain patterns for the $x$-component of the TR field around a point sink in the $xOz$ plane when: (a) $\omega t = \pi/2$; (b) $\omega t = \pi$; (c) $\omega t = \pi/5$. The three columns respectively correspond to spatial field distribution, $k$-domain spectrum on the $k_x$-$k_z$ plane and the line of $k_x=k_z$ (see **Visualization 2**).

Fig. 9 displays the amplitudes on the diagonal of the $k$-space with respect to $\omega t$ from 0 to $2\pi$, where amplitudes along the lines $k = 2k_0/3$ and $k = 4k_0/3$ are chosen to roughly depict the variations for propagating and evanescent waves respectively, as that along the line $k = 0$ maintain zero. It is convincing that the super-resolution should be achieved with a point-like focusing spot except the case when $\omega t = m\pi$ ($m = 0, 1, 2….$), which can be accounted for by the absence of the evanescent waves.

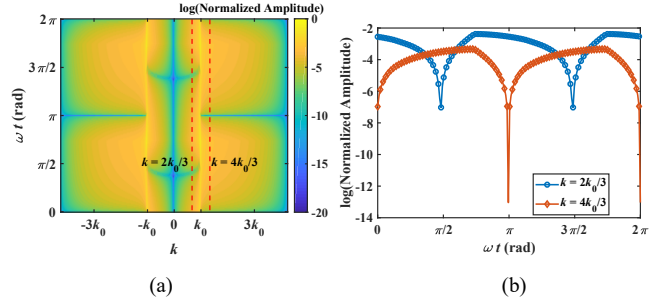

Fig. 9. (a) Normalized spatial spectrum on a logarithmic scale on the diagonal of the $k$-space for the $x$-component of the TR field around a point sink in the $xOz$ plane during a period; (b) the amplitudes along the red dashed lines in (a).

Fig. 10 shows the normalized field distributions and $k$-domain patterns of the $z$-component for $\omega t = \pi/2$, $\omega t = \pi$ and $\omega t = \pi/5$. Different from Fig. 8, the time-varying $k$-domain pattern shows that the propagating waves are always there, while no evanescent waves at $\omega t = m\pi$ ($m = 0, 1, 2….$), leading to the realization of the super-resolution except for $\omega t = m\pi$. What's more, the amplitudes on the line $k_z = 0$ with respect to $\omega t$ from 0 to $2\pi$ drawn in Fig. 11 further corroborate the above argument.

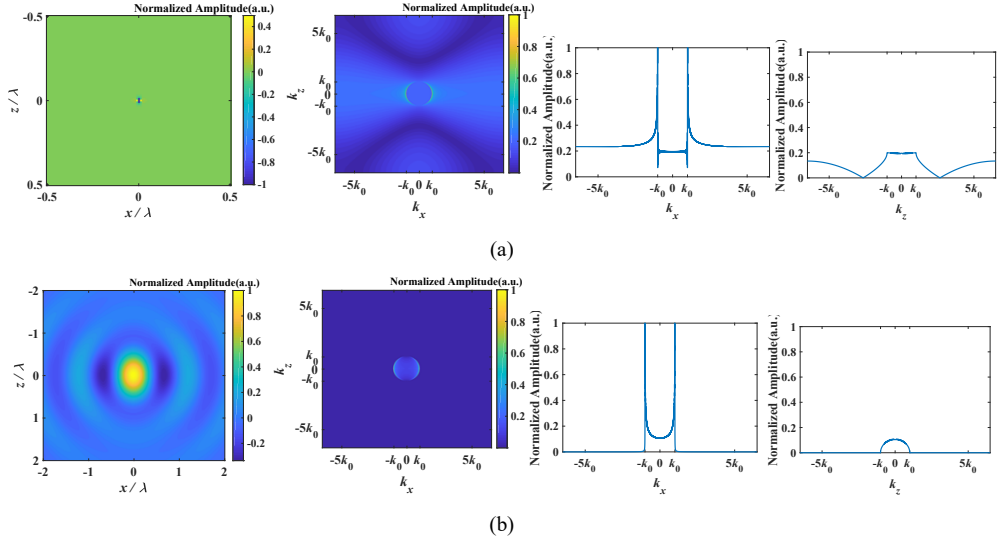

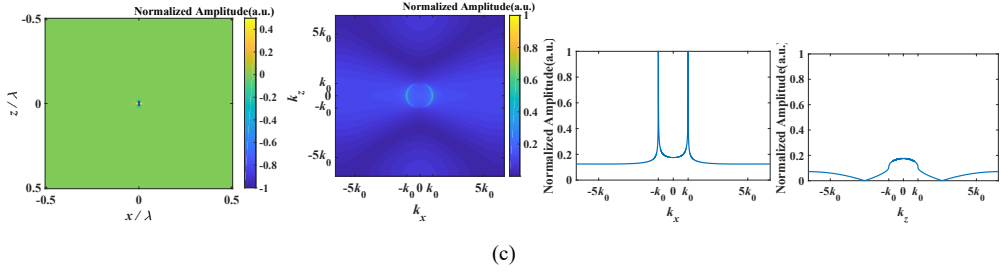

(c)

Fig.10. Normalized field distributions and $k$-domain patterns for the $z$-component of the TR field around a point sink in the $xOz$ plane when: (a) $\omega t = \pi/2$; (b) $\omega t = \pi$; (c) $\omega t = \pi/5$. The four columns respectively correspond to spatial field distribution, $k$-domain spectrum on the $k_x$-$k_z$ plane, the line of $k_z=0$, and the line of $k_x=0$ (see **Visualization 3**).

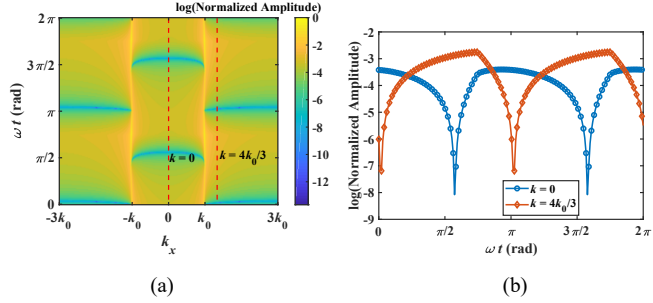

(a)            (b)

Fig.11. (a) Normalized $k$-domain pattern on a logarithmic scale on the line $k_z = 0$ for the $z$-component of the TR field around appoint sink in the $xOz$ plane during a period; (b) the amplitudes along the red dashed lines in (a).

When the $xOy$ plane that the source perpendicular to are investigated, the similar time-varying phenomenon occurs in the $k$-domain patterns, as detailed in Section 2 in **Supplement 1,** the time-varying video for which can be found in **Visualization 4**.

### 3.2 K-domain Modulation with the magnetic current as the initial excitation source

Take the same assumption as Sections 2.2.2 and 2.3.2 that a unit magnetic current source placed at the origin and along the $x$-direction as the initial excitation source, then the instantaneous expressions of the TR field in the $xOy$ plane that the source parallel to corresponding to Eq. (11) and Eq. (13) are calculated as

$$\boldsymbol{E}_{total}^{TR}(R;t) = \mathrm{Re}\left(\left(\frac{j}{2\pi} \cdot \frac{kR\cos kR - \sin kR}{R^3} \cdot y\boldsymbol{e}_z\right) \cdot \exp(j\omega t)\right) \tag{16}$$

$$\boldsymbol{E}_{anti-causal}^{TR}(R;t) = \mathrm{Re}\left(\left(\frac{1}{4\pi} \cdot \frac{\exp(jkR)(jkR-1)}{R^3} \cdot (-y\boldsymbol{e}_z)\right) \cdot \exp(j\omega t)\right) \tag{17}$$

where $\boldsymbol{e}_z$ is the unit vector along the positive $z$-direction.

After the analysis for above cases, a reasonable prediction can be made that the vector total TR field without the point sink in Eq. (16) only consists of propagating waves, while the evanescent waves appear for the TR field around a point sink in Eq. (17), accompanied by the time-varying $k$-domain modulation. The patterns obtained using Eqs. (16) and (17) are shown in Fig. 12 and Fig. 13 respectively, indicating an excellent agreement between the results and our prediction.

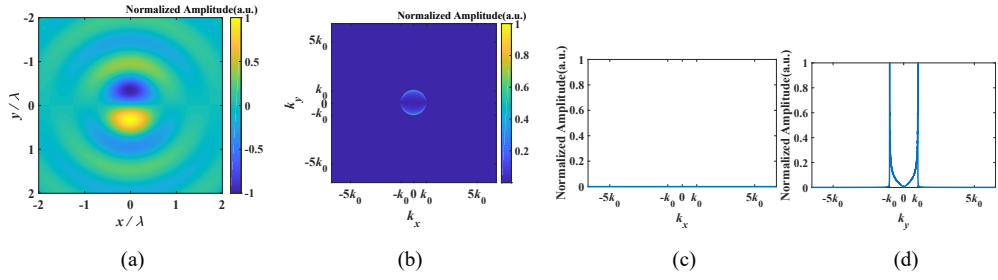

(a)                  (b)                  (c)                  (d)

Fig.12. Normalized field distribution and *k*-domain patterns of the total TR field without the point sink in the *xOy* plane corresponding to unit magnetic current when $\omega t = \pi / 2$: (a)normalized spatial distribution; (b)normalized *k*-domain pattern of (a); normalized *k*-domain patterns along the lines of (c) $k_x = 0$ and (d) $k_y = 0$.

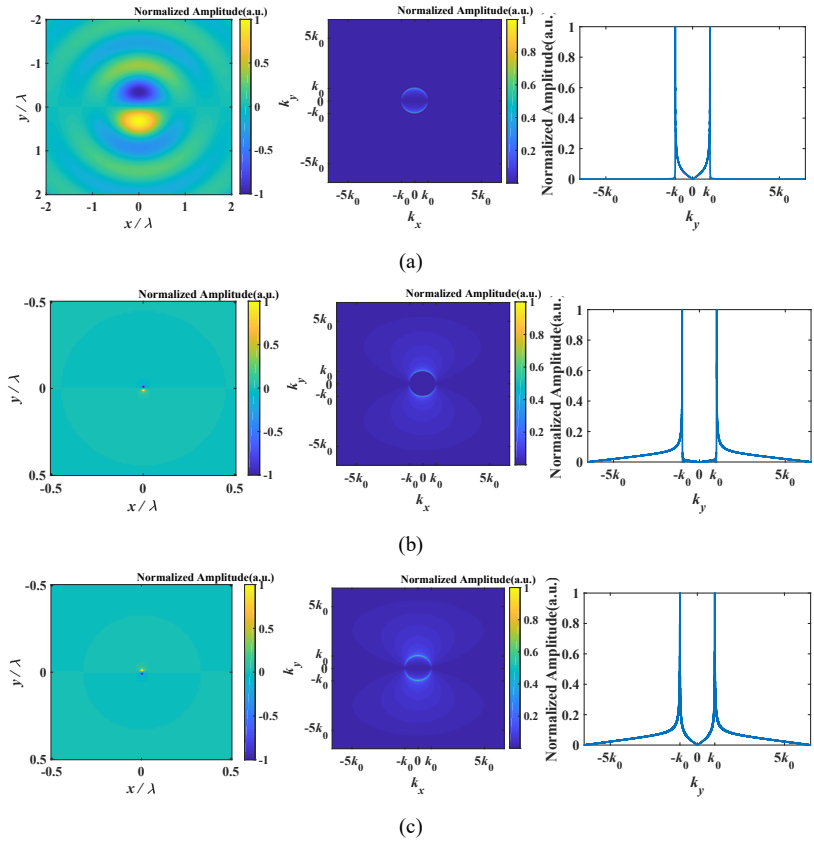

Fig.13. Normalized field distributions and *k*-domain patterns of the TR field around a point sink in the *xOy* plane corresponding to unit magnetic current when: (a) $\omega t = \pi / 2$; (b) $\omega t = \pi$; (c) $\omega t = \pi / 5$. The three columns respectively correspond to spatial field distribution, *k*-domain spectrum on the $k_x$-$k_y$ plane and the line of $k_x$=0 (see **Visualization 5**).

Fig. 13(a) shows a focusing spot constrained by the diffraction limit, and except for the moments $\omega t = m\pi + \pi / 2$ ($m = 0, 1, 2….$) it illustrates, the field indicated by Eq. (17) has a point-like focusing spot, which contributes to the realization of the super resolution.

Fig. 14(a) displays the amplitudes on the line $k_x = 0$ with respect to $\omega t$ from 0 to $2\pi$. The values on the lines $k = 2k_0 / 3$ and $k = 4k_0 / 3$ shown separately in Fig. 14(b) reflects the periodic fluctuations of propagating waves and evanescent waves, respectively.

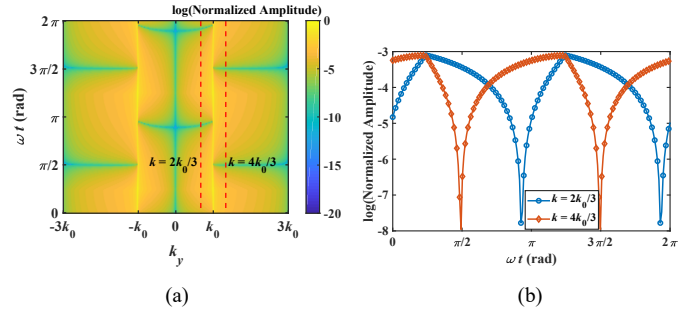

Fig.14. (a)Normalized $k$-domain pattern on a logarithmic scale on the line $k_x = 0$ for the TR field around a point sink in the *xOy* plane during a period; (b)the amplitudes along the red dashed lines in (a).

The time-varying $k$-domain patterns in the *yOz* plane that the source perpendicular to are similar to that in the *xOy* plane, and the detailed analyses can be found in Section 3 in **Supplement 1**. The time-varying video in this case is shown as **Visualization 6**.

In summary, regardless of the scalar TR field or the vector TR field with the electric or magnetic current as the initial excitation source, the evanescent waves can always be observed in the far field if a point sink is set at the initial source point, leaving only the converging waves represented by the anti-causal Green's function. That is, the $k$-domain modulation from the evanescent waves with large transverse wavenumber to propagating waves with small transverse wavenumber contributes to the realization of the super resolution, and this $k$-domain modulation is observed time-varying.

## 4. Discussions and Conclusion

The time-varying $k$-domain modulation based on TR technique is investigated in this paper, which contributes an efficient approach to the optical super-resolution imaging and focusing beyond the classical diffraction limit without the need for structurally engineered materials. According to the proposed TR field decomposition theory, the TR field can be decomposed into diverging waves and converging waves, characterized by the causal Green's function and the anti-causal Green's function, respectively. It is revealed that the super-resolution imaging can be achieved if the isolated anti-causal Green's function-based field is formed by putting a point sink at the initial source point during the process of TR focusing. The reason behind this super-resolution property lies in the $k$-domain modulation, that is, the evanescent waves with large transverse wavenumbers are spontaneously converted into propagating waves with small transverse wavenumbers, transferring the subwavelength features into the far field. Furthermore, the novel phenomenon of time-varying $k$-domain modulation around a point sink is demonstrated when TR source is either a current element or a magnetic current element. It is expected that the proposed theory can be used for microwave/optical super-resolution focusing and imaging via TR systems, facilitating upcoming applications such as high spatial-resolution wireless communication, directive high-power energy transfer, and distributed oncotherapy.

**Funding.** National Natural Science Foundation of China (62171081 and U2341207), the Natural Science Foundation of Sichuan Province (2022NSFSC0039), and Aeronautical Science Foundation of China (2023Z062080002).

**Acknowledgments.** Authors thank Jiang Xiong for useful discussions.

**Disclosures.** The authors declare no conflicts of interest.

**Data availability.** Data and codes underlying the results presented in this paper can be obtained from the authors upon reasonable request.

**Supplemental document.** See **Supplement 1** for supporting content.